# π-Conjugation and conformation in a semiconducting polymer: small angle x-ray scattering study

Paramita Kar Choudhury[*], Debjani Bagchi, and Reghu Menon

*Department of Physics, Indian Institute of Science, Bangalore 560012, India*

Small angle X-ray scattering (SAXS) in poly[2-methoxy-5-(2'–ethyl-hexyloxy)-1,4-phenylene vinylene] (MEH-PPV) solution has shown the important role of π-electron conjugation in controlling the chain conformation and assembly. By increasing the extent of conjugation from 30 to 100 %, the persistence length ($l_p$) increases from 20 to 66 Å. Moreover, a pronounced second peak in the pair distribution function has been observed in fully conjugated chain, at larger length scales. This feature indicates that the chain segments tend to self-assemble as the conjugation along the chain increases. Xylene enhances the rigidity of PPV backbone to yield extended structures, while tetrahydrofuran solvates the side groups to form compact coils in which the $l_p$ is much shorter.

*Email: s_paramita@physics.iisc.ernet.in

Earlier photophysical studies in semiconducting polymers like poly[2-methoxy-5-(2' – ethyl-hexyloxy)-1,4-phenylene vinylene] (MEH-PPV) have shown that the nanoscale self-organization of chains affects the photoluminescence (PL), electroluminescence and fluorescence properties [1, 2, 3]. For example, PL in PPV derivatives shows prominent red-shift due to the enhanced energy transfer processes as the π-electrons delocalization increases [4]. This indicates that as the conjugation length increases, the π-π interactions can give rise to nanoscale aggregates; as a result the PL spectra undergoes a red shift. Furthermore, the torsional angle between the benzene rings in polymer backbone tunes the conjugation length [5], and also the organization of chain segments via the interchain interactions that in turn modify the fluorescence spectrum and PL quantum yield. In fluorescence studies, the quantum yield is observed to decrease when the interchain energy transfer increases. Moreover, the small-angle neutron scattering (SANS) study in dilute solutions of MEH-PPV has shown the aggregation of chain segments to form nanoscale disc-like domains [6]. Hence more detailed studies regarding how conjugation and stiffness of chains affect the nanoscale organization of chain segments is required. An understanding of the role of chain conformation is important in device fabrication since the memory of the structural information in solution is retained in the solid films [7, 8].

In this respect, the small angle X-ray scattering (SAXS) studies in polymeric systems have shown that the local nanoscale morphology at various length scales can be probed to show the correlation among conformation and assembly of chains [9, 10]. Although the SAXS studies in conjugated polymer systems are rather few, the preliminary work in Poly(3,4-ethylenedioxythiophene)-poly(styrenesulfonate) [PEDOT-PSS] has shown how the solvent modifies the chain conformation [11]. The SANS experiment in MEH-PPV has shown the role of solvents in nanoscale aggregation and also the formation of macroscopic nematic phase. Nevertheless how the conjugation length affects the organization of chains is not being understood yet. The earlier dynamic light scattering (DLS) work in PPV derivatives has shown how the variation in side group can influence the conjugation length, aggregation of chains and PL spectra [12]. This clearly shows the significance of the correlation among conformational and physical properties, which requires more detailed SAXS investigations. In the present study the conjugation length in MEH-PPV samples are controlled by selective thermal elimination, as already inferred from the absorption peaks [4]; hence the role of conjugation length has been investigated in the SAXS data.

SAXS measurements were carried out using Bruker Nanostar equipped with a rotating anode source and three-pinhole collimation. A position sensitive 2D detector with 100 μm resolution was used to record the scattered intensity. The scattered intensity $I(q)$ is plotted as a function of the momentum transfer vector $q = 4\pi \sin\theta / \lambda$, where $\lambda$ is the wavelength of the X-rays (Cu-kα radiation, 1.54 Å), and $\theta$ is half the scattering angle. The $q$-range is 0.008 Å$^{-1}$ < $q$ < 0.3 Å$^{-1}$. The raw data was normalized for transmission coefficient, capillary width and exposure time; and also the incoherent scatterings due to solvent were subtracted in the data analysis. The MEH-PPV samples in this study are of three types: fully conjugated (M-PPV100, ~ 100%), intermediate conjugation (M-PPV70, ~ 70 %) and less conjugated (M-PPV30, ~30%); the corresponding absorption peaks are at 508, 450, and 425 nm, respectively [4]. The dilute solutions (1% by weight) of these samples were prepared in xylene. These solutions were filled in Mark glass capillary of two mm diameter and sealed. Interestingly, the samples can be distinguished by their color (M-PPV100 is red, M-PPV70 is orange and M-PPV30 is yellow), as shown in the inset of Figure 1. Furthermore, how the solvent affects the SAXS data for M-PPV100 is being studied by comparing the results in xylene and tetrahydrofuran (THF).

Figure 1 shows the log-log plot of $I(q)$ vs. $q$. The slopes of $I(q)$ are $-1$ for $q > 0.1$ Å$^{-1}$. The slopes of $I(q)$ are $-2$ for $q < 0.1$ Å$^{-1}$, in case of M-PPV100 and M-PPV70, and the shift in slopes occur at $q \sim 0.03$ and $0.05$ Å$^{-1}$, respectively. However, in M-PPV30 this slope change at $0.1$ Å$^{-1}$ is hardly noticeable. The cut-off for this change in slope shifts to lower q values as the extent of conjugation increases. The solutions are rather dilute (1%) so that the scattering is due to the form factor of well-separated chains. It is known that the slope -1 corresponds to the rigid-rod structure of chain segments [10]. This is in accordance with the fact that conjugation can make the segments more rigid. However, at larger length scale ($q \leq 0.03$ Å$^{-1}$) the presence of flexible coil like features can be observed. In M-PPV70 and M-PPV100 the slope change is rather abrupt as $q$ decreases; and this is attributed to the formation of aggregates, in which the rod-like segments assist self-assembly. This indicates that the extent of conjugation enhances the organization of chains, and this is perceptible in other physical properties like charge transport. These features are further analyzed to obtain more detailed structural parameters.

Figure 2 shows the intensity profile of M-PPV100, M-PPV70 and M-PPV30 in xylene. The scattering profile follows a Debye function that decreases as $q^{-2}$ at small $q$ values, which reflects the coiled nature; and then shows a $q^{-1}$ behavior at larger $q$ as in rigid rods. The degree of rigidity as a measure of the curvature of chain can be estimated from persistence length ($l_p$). The critical value $q^*$ at the crossover between these two regions is related to the persistence length by the equation[13]: $q^* l_p = 1.91$. The $q^*$ values for M-PPV100, M-PPV70, and M-PPV30 are 0.03, 0.05, and 0.1 Å$^{-1}$, and the estimated values ($\pm 5\%$) of $l_p$ are 65, 41, and 20 Å, respectively. It is interesting to note that $l_p$ for M-PPV100 is three times larger than that of M-PPV30, in accordance with the extent of conjugation. This implies that the fully conjugated chains are more rigid compared to the less conjugated ones.

The fits in figure 2 follow the modified *worm-like chain* (WLC) model [14, 15, 16]. It deals with the local stiffness of chain segments at lower $q$ [parameterized by Kuhn length ($b$)], and also by the Gaussian coiled structure at higher $q$ values; hence the model can be used to fit the data in the entire $q$-range, that takes into account the excluded volume effect [17]. The solid lines are fit to Eqn. 1 for the prescribed model [14]; and the fit parameters are shown in Table I.

$$S_{exv}(q) = w(qR_g) S_D(q,L,b) + [1 - w(qR_g)][C_1(qR_g)^{-1/\nu} + C_2(qR_g)^{-2/\nu} + C_3(qR_g)^{-3/\nu}] \quad (1)$$

where $S_D = S_{Debye}(q,L,b) = 2[exp(-u) + u - 1]/u^2$ and $u = q^2 R_g^2$ \quad (2)

The function $w(qR_g)$ is a crossover function:

$w(x) = [1 + tanh((x - C_4)/C_5)]$ with $x = qR_g$ \quad (3)

The radius of gyration, $R_g$ is calculated by:

$<R_g^2> = (Lb/6)[1 - (3/2n_b) + 3/2n_b^2 - 3/4n_b^3(1 - exp(-2n_b))]$ \quad (4)

where $n_b$ (=$L/b$) is the number of Kuhn segments; $L$ is the chain contour length. The Kuhn length '$b$' is related to $l_p$ by $2l_p = b$. The value of L is calculated from $L = l_0 (M/M_0)$ where $l_0 = 6.7$ Å is the length of a monomer [18], and $M$ and $M_0$ are the molecular weights [4] of polymer $M = 250000$ and monomer $M_0 = 276$. Using this relation, $L= 6068.8$ Å. These values of $l_p$ from the fits can be compared to the ones obtained from $q^*$ $l_p= 1.91$, and it agrees well. In this analysis, although different structural length scales (from Kuhn segment to radius of gyration) are represented by different decades of $q$ in the intensity profile, the modified WLC Model can fit the entire range of $I$ vs. $q$, as shown in figure 2. The features present in figures 1 and 2 indicate that the chains are represented by a combination of both flexible and rigid segments, as in the modified WLC model. The values of $R_g$ and $l_p$ for M-PPV100 (Table 1) can be compared to those from DLS (520 and 60 Å) [12]; and observed to be quite consistent. Since the values of $l_p$ for M-PPV100 is nearly three times larger than that of M-PPV30, the enhanced π-π interaction in former favors the self-organization of chains, which is rather weak in latter. This type of self-assembly of the chain segments is shown in the schematic diagram in figure 2. In fact this feature in the solution state is retained even in the solid films that in turn play a major role in the device characteristics [19].

The above data analysis is further confirmed from the pair distribution function (PDF) analysis, since the structural information in earlier analysis (in Fourier space) has a correspondence with the conformational features in real space. The pair (distance) distribution function $p(r)$ [the probability distribution of the vectors between pairs of scatterers] is calculated by inverse Fourier transform of the scattered intensity I(q), using the algorithm GNOM [21]:

$$P(r) = \frac{1}{2\pi^2} \int_{q\min}^{q\max} I(q) qr \sin(qr) dq \qquad (5)$$

The average size of the chain $<R_g>$ can be calculated from the second moment of $p(r)$:

$$\langle R_g \rangle^2 = \frac{\int_0^{r_{max}} r^2 p(r)dr}{2\int_0^{r_{max}} p(r)dr} \tag{6}$$

The *p(r) vs. r* (distance between a pair of scatters) data for the three samples in xylene are shown in figure 3. All the three samples show two peaks that correspond to two types of correlations at different length scales: first one for a local short range and the second for long-range order [11]. The first peak corresponds to intra-chain correlations among the segments, and the corresponding maximum in r gives the most probable distance for the scatterers in the chain, i.e. the chain contour length, which consists of several Kuhn segments. The second peak accounts for inter-chain correlation and the corresponding maximum in r indicates the distance to which the chains can be correlated to each other [21]. It is interesting to note that the second peak is quite pronounced for M-PPV100, showing that the organization of chain segments (due to enhanced π-π interactions) is considerably more with respect to the less conjugated samples. Moreover, the shifting of the second peak towards larger r, as the conjugation increase, suggests that the size of ordered domains increases. This implies that the formation of a nematic phase due to the long range assembly of segments is quite possible, as also been reported from the SANS data. This is found to be in agreement with the features observed in optical properties too. Furthermore, the absorption spectra are also observed to undergo a red-shift as the conjugation length increases [4]. Although these spectral features are quite evident, how the conformation of chains play a role in this is not yet understood. Nevertheless, the SAXS data show that the second peak in PDF is due to the assembly of stiffened chain segments that corroborates with the red-shift in absorption and PL. The values of $R_g$ and $l_p$ from PDF analysis are given in Table II. These values agree very well with the parameters obtained from the fit to the modified WLC model, as in Table 1. Apart from these conjugation dependent properties, the optical properties of MEH-PPV have shown interesting solvent dependent behavior [22], which is yet to be explored in SAXS measurements.

Earlier work in MEH-PPV has shown that the optical properties in PL and absorption spectra can be tuned by the polarity of solvent [5, 22]. The SAXS data for M-PPV100, in dilute solutions of xylene (non-polar) and THF (polar), are shown in figure 4. The data in THF solution hardly show any significant variation in the slope while compared to that in xylene. It is well known that the nature of solvent strongly modifies the conformation of chains. It has been observed that the tendency to form aggregates increases in non-polar solvents than that in polar solvents. Non-aromatic solvents like THF prefer solvating the side groups of MEH-PPV, while solvents like xylene solvate the polymer backbone that give rise to different conformations [5]. The increase in slope at lower $q$, for xylene, is not observed in the THF solution. This implies that the long-range assembly of chain segments does hardly occur in THF solution. This is due to the fact that the solvation layer of THF molecules around the MEH side groups, and the relatively flexible backbone, facilitates the formation of compact coils in which the π–electrons are more localized as already been observed in the optical studies [22]. On the other hand, the preferential interaction of conjugated backbone with solvents like xylene enhances the planar conformation of segments that favor the formation of extended structures in which the π–electrons are more delocalized; that also contributes to the long range self-organization. This is further quantified by comparing the values of persistence length in xylene and THF solutions, as 66 Å and 49 Å respectively; and the $q^*$ values are ~ 0.03 and 0.038 Å$^{-1}$. The SAXS data in M-PPV100 in THF may resemble with that for M-PPV30 in xylene. Since the THF solvation of MEH side groups can increase the torsional angle between the rings in PPV chain, as a result the effective conjugation length decreases, and its implication has been observed in the optical properties [5]. For example, the absorption peak shows a red-shift as the conjugation length increases; also the PL drifts to longer conjugation lengths before the radiative relaxation decay [23, 24]. Interestingly, a similar decrease in $l_p$ has been reported in a biopolymer due to the compact conformation [25]. These results are quite relevant and should be taken into account in device fabrication, since the interchain interactions and aggregation depends on the choice of solvent, and the chain conformation in the solution phase tunes the charge-transfer process in the solid films [26].

In summary, the SAXS data show modifications in chain conformation and self-assembly as the extent of conjugation and π–electron delocalization in MEH-PPV samples varies. The persistence length increases by a factor of three as the conjugation increases from 30 to 100 %. The presence of a prominent second peak in M-PPV100, at larger length scales, shows the presence of long-range order. This feature can be tuned by selecting the nature of solvent, as has been observed in case of xylene and THF.

We thank Prof. S. Ramakrishnan for samples, and Prof. K. Guruswamy and K. Sharma for facilitating the experiments. We thank DST for funds.

Figure Captions:

Fig. 1. log I vs. log q for MEH-PPV samples in xylene. The photo shows the visible change in colour due to the change in conjugation. The chemical structure of the monomer is shown in the inset.

Fig. 2. Intensity vs. q for M-PPV100 (□), M-PPV70 (△), and M-PPV30 (○) in xylene. Solid lines are fit to Eqn 1. The inset shows a schematic diagram of various conformation and organization of chain segments.

Fig. 3. Pair Distribution Function vs. distance (r). The baselines are shifted for clarity. The first and second peaks show the correlation among intra & inter-chain segments.

Fig. 4. log I vs. log q for M-PPV100 in xylene and THF. The change in slope is prominent in xylene, which is absent in THF.

Table Captions:

Table I. Values of parameters $C_1$, $C_2$, $C_3$, $C_4$, $C_5$, $R_g$, and $l_p$, as obtained from fits to Eqn 1

Table II. Values of $R_g$ and $l_p$ from the PDF analysis.

Figures

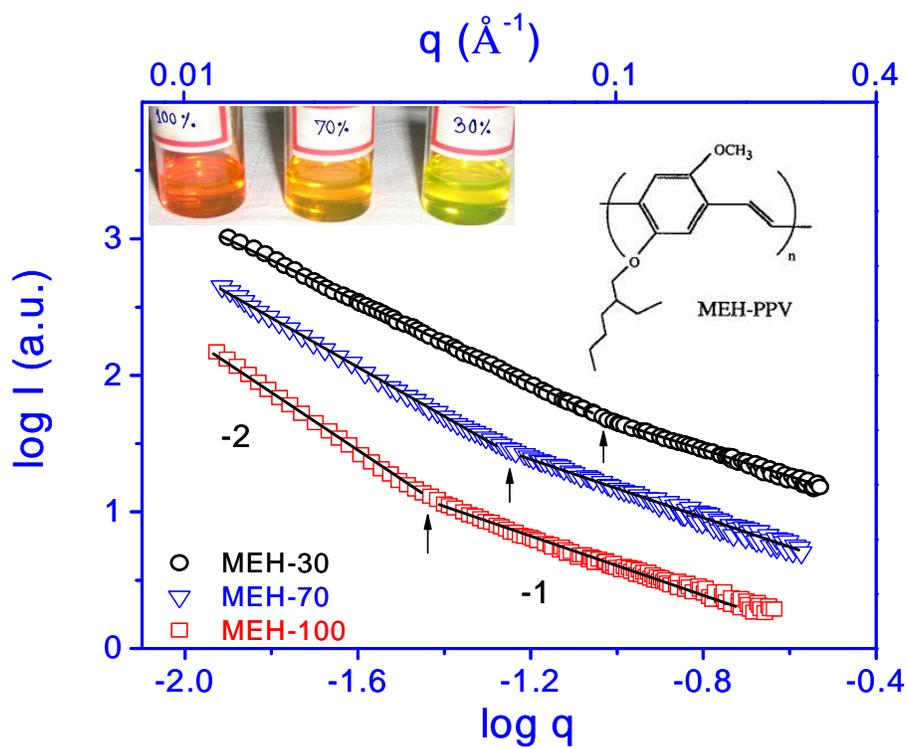

Fig. 1. log I vs. log q for MEH-PPV samples in xylene. Baselines are shifted for clarity. The photo shows the visible change in colour due to the change in conjugation. The chemical structure of the monomer is shown in the inset.

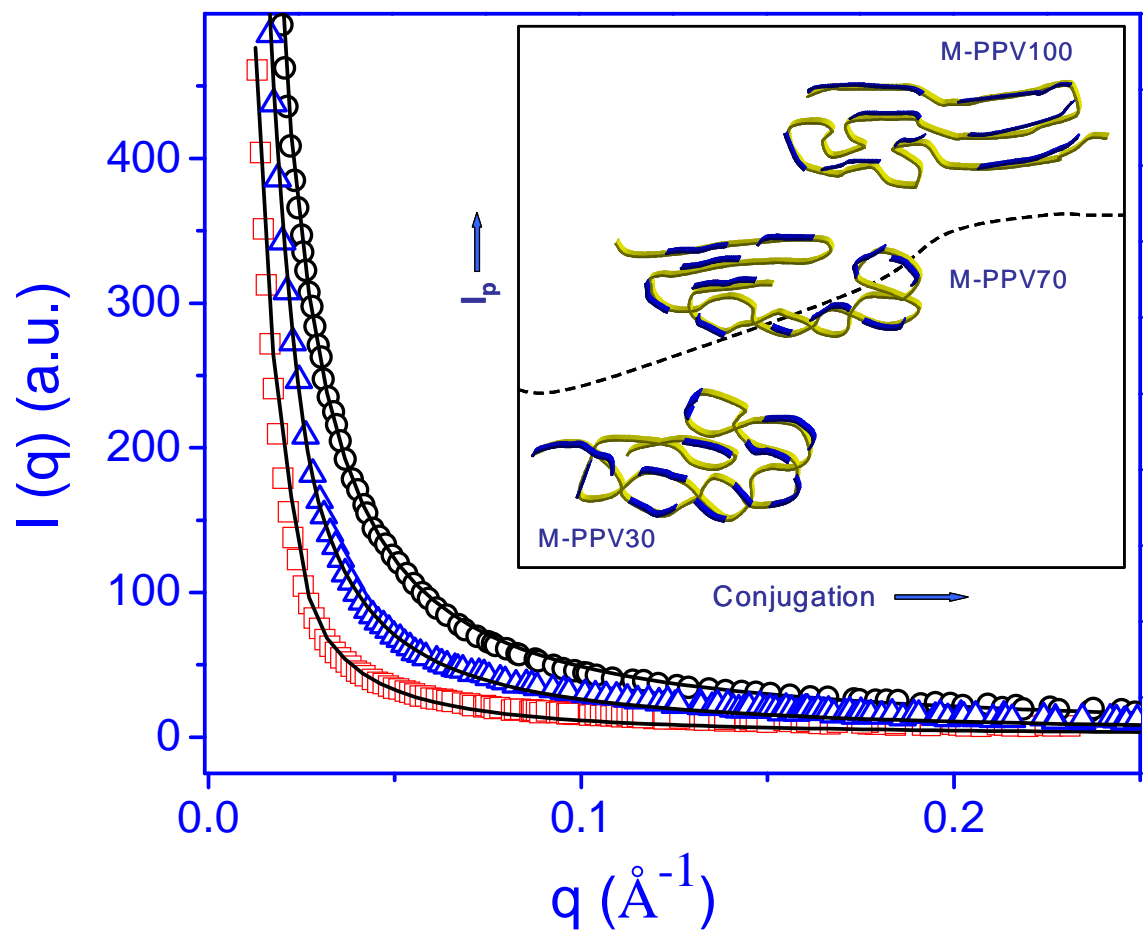

Fig. 2. Intensity vs. q for M-PPV100 (□), M-PPV70 (△), and M-PPV30 (○) in xylene. Solid lines are fit to Eqn 1. The inset shows a schematic diagram of various conformation and organization of chain segments.

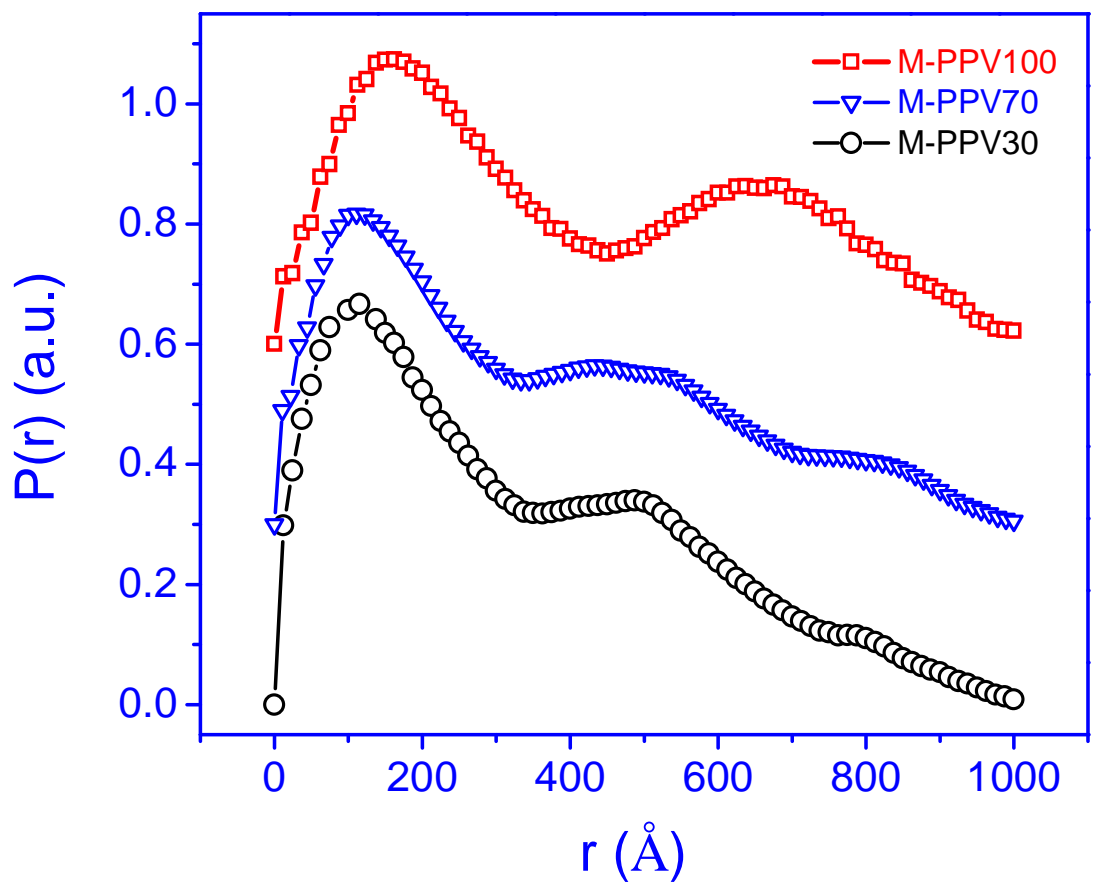

Fig. 3. Pair Distribution Function vs. distance (r). The baselines are shifted for clarity. The first and second peaks show the correlation among intra & inter-chain segments.

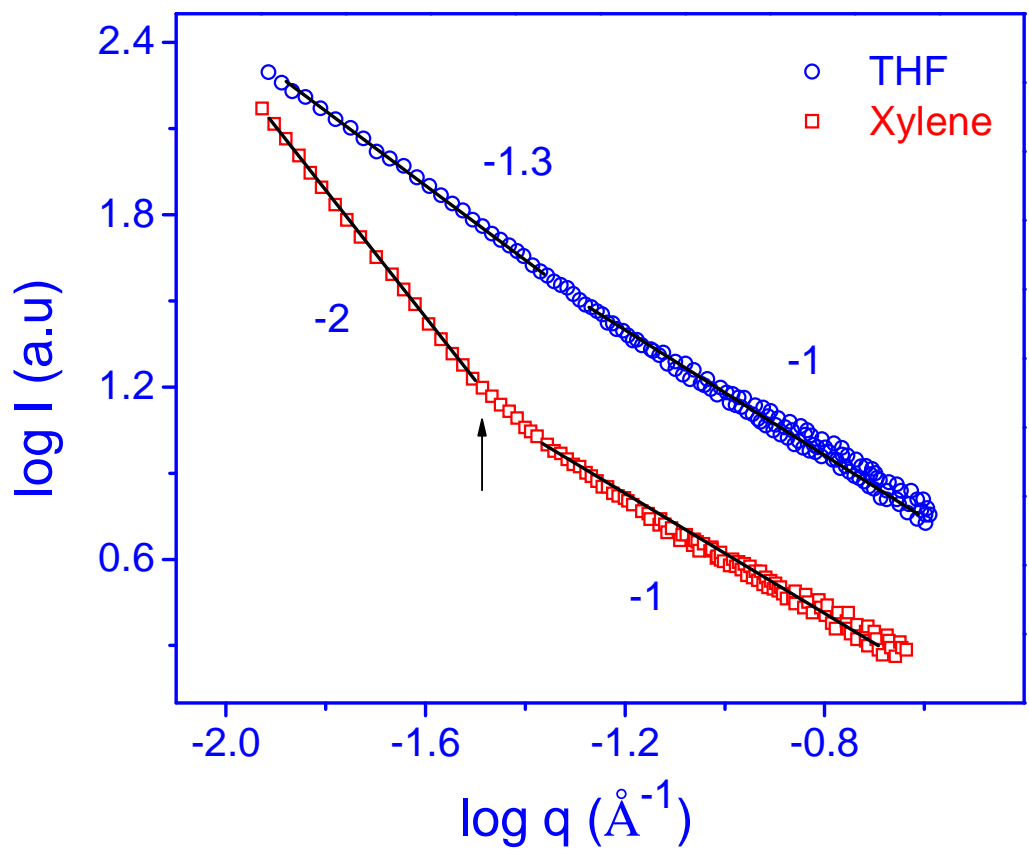

Fig. 4. log I vs. log q for M-PPV100 in xylene and THF. The change in slope is prominent in xylene, which is absent in THF.

Tables

| Sample | $C_1$ | $C_2$ | $C_3$ | $C_4$ | $C_5$ | $R_g$ (Å) | $l_p$ (Å) |
|---|---|---|---|---|---|---|---|
| MEH-100 | 583 | 14346 | -1.65 | 9 | 0.15 | 360 ± 10.8 | 66.18 ± 2.8 |
| MEH-70 | 1070 | 16344 | -1.65 | 7 | 0.07 | 290 ± 7.6 | 42.45 ± 1.1 |
| MEH-30 | 1490.2 | 9743.5 | -1.65 | 2.32 | 0.003 | 204 ± 4.8 | 20.78 ± 0.9 |

Table I. Values of parameters $C_1$, $C_2$, $C_3$, $C_4$, $C_5$, $R_g$, and $l_p$, as obtained from the fit to Eqn 1

| Sample | $<R_g>$ (Å) | $l_p$ (Å) |
|---|---|---|
| MEH-100 | 336.25 ± 13.5 | 57.4 ± 3.3 |
| MEH-70 | 297 ± 10.6 | 44.5 ± 1.44 |
| MEH-30 | 209 ± 8.2 | 21.87 ± 0.67 |

Table II. Values of $R_g$ and $l_p$ from the PDF analysis.